\documentclass[conference]{IEEEtran}
\IEEEoverridecommandlockouts
\usepackage{cite}
\usepackage{amsmath,amssymb,amsfonts}
\usepackage{graphicx}
\usepackage{textcomp}
\usepackage{xcolor}
\usepackage{amsmath, amssymb}
\usepackage{algorithm}
\usepackage{algpseudocode}
\usepackage{physics}
\usepackage{float} 
\usepackage{subcaption}
\usepackage{booktabs}
\usepackage{multirow}
\usepackage{wrapfig} 
\usepackage{array}    
\usepackage{caption}  
\usepackage{adjustbox}
\usepackage{soul}
\usepackage{makecell}
\usepackage{fancyhdr}
\def\BibTeX{{\rm B\kern-.05em{\sc i\kern-.025em b}\kern-.08em
    T\kern-.1667em\lower.7ex\hbox{E}\kern-.125emX}}
    
\begin{document}

\title{Empirical Evaluation of Cross-Carrier MCPTT \& OTT MCX Interoperability  in High-Density Environments
}

\author{
  \IEEEauthorblockN{
    Eman Hammad\IEEEauthorrefmark{3} \IEEEauthorrefmark{4},
    Derek Ladd\IEEEauthorrefmark{1}, 
    Sridhar Kowdley\IEEEauthorrefmark{2}, 
    Walt Magnussen\IEEEauthorrefmark{1}, 
    and Michael Fox\IEEEauthorrefmark{1} 
  }
  \IEEEauthorblockA{
    \IEEEauthorrefmark{4}Texas A\&M University, College Station, TX, USA \\
    \IEEEauthorrefmark{1}Center for Applied Communication and Networks (CACN) \\ 
    \IEEEauthorrefmark{2}Department of Homeland Security (DHS), Washington, DC, USA \\
    \IEEEauthorrefmark{3}iSTAR Laboratory, Department of Engineering Technology and Industrial Distribution \\
    Emails: \IEEEauthorrefmark{3}eman.hammad@tamu.edu, \{\IEEEauthorrefmark{1}dladd, wmagnussen, mfox\}@tamus.edu, 
    \IEEEauthorrefmark{2}svkowdley@gmail.com 
  }
}

\maketitle
\vspace{-15pt}

\begin{abstract}

Deploying broadband Mission-Critical Push-To-Talk (MCPTT) services over shared commercial infrastructures introduces resource contention during multi-agency responses in mass-crowd events. This study evaluates cross-carrier interoperability and standard versus prioritized quality of service (QoS) frameworks under real-world saturation constraints. We design an empirical multi-carrier field experiment utilizing twelve identical smartphones deployed across multiple physical sectors inside Texas A\&M University's Kyle Field during a football game with 105,000+ attendees. Automated voice calls were monitored using Perceptual Objective Listening Quality Analysis (POLQA), packet delivery metrics, and connection rates. The results reveal that voice path failure is isolated to network infrastructure bottlenecks rather than device hardware limitations. Specifically, we identify a sharp, non-linear network failure model where transport-layer jitter exceeding a critical threshold de-jitter buffer underflows, causing structural audio degradation. Priority-managed channels effectively bypass this congestion. This study helps establish an  operational insight for emergency planners to mandate network infrastructure, end-to-end network slicing and dedicated resource provisioning capable of keeping transport-layer jitter below the critical failure boundary.

\end{abstract}


\fancyhead[C]{This manuscript was accepted at 2026 IEEE World-Forum on Public Safety Technologies (IEEE WF-PST 2026).}
\section{Introduction}
\label{sec:intro}

\IEEEPARstart{M}{ission} Critical Services (MCX) over broadband networks are replacing voice-centric public safety architectures with high-throughput, data-rich ecosystems. Achieving deterministic reliability under extreme operating constraints remains a core engineering challenge as agencies shift toward flexible, software-defined network profiles. Currently, public safety coordination relied on Land Mobile Radio (LMR) networks, which is designed to provide highly reliable mission critical voice communications using narrow channel bandwidth which does not support video, real-time biometrics, and complex data telemetry. \cite{favraud2016towards}. To overcome these limitations, the sector is transitioning to broadband infrastructures like FirstNet and commercial priority layers \cite{baldor2018public}. This introduces LTE and 5G technologies into the mission-critical domain, where Mission-Critical Push-To-Talk (MCPTT) frameworks deliver LMR-like voice performance over shared commercial Radio Access Networks (RAN) \cite{3gpp_ts22179}. However unlike LMR which are exclusively designed for public safety, broadband networks share the network with commercial traffic creating resource challenges  under high network load.

A critical requirement for modern public safety broadband is cross-carrier reliability. During multi-jurisdictional emergencies or mutual aid events, responders from different agencies must coordinate seamlessly regardless of whether they operate on the same carriers or not.\cite{ferrus2013lte}. This introduces potential interoperability bottlenecks at network boundaries; while individual carriers optimize internal radio resources, cross-carrier routing requires the seamless interworking of priority mapping policies, distinct core gateways, and varying transport-layer queuing behaviors. Excessive latency or packet delay variations at these boundaries can rapidly degrade the end-to-end voice path, making it essential to evaluate prioritized interoperable profiles across multiple distinct carrier infrastructures simultaneously.

The current market offerings provide two types of push-to-talk applications: Over-the-Top (OTT) applications, which are often proprietary, non-standard solutions that lack cross-organizational interoperability, and integrated core services, specifically 3GPP Mission-Critical Push-to-Talk (MCPTT). While OTT applications generally operate as \textit{best-effort} services without underlying resource pre-emption or bearer guarantees, native MCPTT utilizes standardized interfaces to signal the IP Multimedia Subsystem (IMS) core, which in turn directs the 4G/5G core to support priority and preemption. Critical communications need to work during disasters, emergencies, or during heavy traffic conditions such as mass spectator events overloading networks. A packed football stadium serves as an ideal situation for public safety communications experiments, as tens of thousands of competing consumer devices cause uplink and downlink resource block starvation across macro networks and localized Distributed Antenna Systems (DAS) \cite{gomez2017empirical}. This extreme load triggers localized Media Access Control (MAC) layer scheduling delays and compounding backhaul queue depths, causing severe packet delivery disruptions for standard Over-The-Top (OTT) applications. Under these conditions, it remains unverified whether standard Quality of Service (QoS) tiering can effectively isolate cross-carrier MCPTT streams, or if multi-network boundaries introduce unique failure states. A high-capacity college football stadium serves as an ideal field-test environment, replicating the localized infrastructure strain and data starvation of major multi-agency incidents without real-world hazard. To focus findings strictly on fundamental infrastructure mechanics, commercial carrier identities and handset models are intentionally anonymized.

\begin{table*}[t]
\centering
\caption{MCX and MCPTT Regional Initiatives, Projects, and Standard Mechanisms}
\label{tab:mcx_regional_standards}
\scriptsize
\begin{tabular}{lllll}
\toprule
\textbf{Mechanism / Region} & \textbf{Technical Standards} & \textbf{Operational Guidelines} & \textbf{North American Implementation} & \textbf{European Implementation} \\ \midrule
\textbf{MCX Core Services} & 3GPP TS 23.280, TS 22.179 & SAFECOM Continuum, NIFOG & FirstNet, ATIS Adoption & ESN (UK), RRF (France), ETSI ENs \\ 
\textbf{Interworking Function} & 3GPP TS 23.283, TS 100 392 & SCIP Methodologies & P25-LTE Interworking & TETRA-MCX Interworking \\ 
\textbf{Plugtests \& Validation} & ETSI TS 103 564 & ETSI Plugtests Framework & Vendor \& Operator Participation & ETSI Plugtests, Vendor Validation \\ 
\textbf{Certification} & GCF, TCCA IOP & TCCA IOP Certification & GCF, FirstNet Device Approval & GCF, TCCA IOP, National Certs \\ \bottomrule
\end{tabular}
\end{table*}

achieving high availability and inter operable communications is essential, as communication outages directly risk human life. Recent reports attribute localized Mission Critical Services (MCX) disruptions to vendor incompatibilities, non-standard interfaces, and complex legacy Land Mobile Radio (LMR) coexistence \cite{plugtest25, NISTMCRoundtable25}. Non-standardized routing and incomplete 3GPP implementations routinely cause failed group calls and systemic communication breakdowns during joint emergencies \cite{plugtest25}, whereas open, standards-based systems yield seamless cross-agency collaboration. The scale of this challenge was quantified during the 9th ETSI MCX Plugtests (February 2025), where 16.7\% of 174 executed interoperability test cases resulted in functional failures \cite{plugtest25}. These anomalies emerged from structural signaling and media plane incompatibilities, legacy LMR interworking failures (e.g., P25, TETRA), unexpected off-network device behavior, and standards ambiguities \cite{plugtest25}. Empirically characterizing these failure boundaries is vital to ensure critical voice pathways remain uncompromised during dynamic, real-world crises.

This paper presents an empirical cross-carrier field evaluation of prioritized versus standard Over-The-Top (OTT) traffic profiles over multiple active commercial networks inside Texas A\&M University's Kyle Field during a peak mass-spectator event (with 105,000+ attendees). Utilizing Perceptual Objective Listening Quality Analysis (POLQA) Mean Opinion Scores (MOS), transport-layer network jitter, and call success rates, we identify performance gaps across carriers \cite{itu_p863}. The analysis identifies a strict, non-linear network failure threshold demonstrating how transport-layer packet disruptions trigger voice path failures. Finally, we prove these failure boundaries are mainly impacted by network infrastructure, establishing an empirical basis for advanced network resource provisioning models such as dynamic slicing, which provides a virtualized architectural framework that partitions shared physical infrastructure into dedicated, end-to-end logical networks with isolated resource guarantees, to safeguard emergency communications in high-density environments~\cite{etsi3gpp-arch-rel17, chowdhury2025framework}.

\section{Technical Background and Interoperability Frameworks}
\label{sec:background}

The operational paradigm of mission-critical communication is governed by international technical standards and regional regulatory architectures. Globally, the baseline for voice, data, and video interoperability is dictated by the 3GPP Mission Critical Services (MCX) specifications alongside legacy TETRA, Tetrapol, and P25 LMR standards \cite{3gpp_ts23280, tetrapol_std, p25_std}. In North America, broadband capability is anchored by the TIA P25 standard alongside 3GPP-compliant national networks like FirstNet and the regional adoption of these protocols by ATIS~\cite{firstnet_ref}. In Europe, ETSI and the TCCA spearhead systematic migrations from TETRA/Tetrapol toward standardized broadband solutions via formal Plugtests to verify multi-vendor device interoperability \cite{plugtest25, etsi_ts103564}. Beyond technical standards, operational frameworks manage joint-agency execution. In the United States, CISA coordinates this via the SAFECOM Interoperability Continuum, NIFOG, and SCIP methodologies \cite{cisa_safecom, InteroContinuum}, while European validation relies on the TCCA IOP Certification process and GCF approvals. Aligning these disparate regional policies requires deep network-layer optimization of the Interworking Function (IWF) to map distinct Quality of Service (QoS) classes across multi-carrier boundaries.

Historically, public safety personnel relied on the low-latency, high-availability Push-to-Talk (PTT) capabilities of legacy LMR architectures like P25 and TETRA \cite{p25_std}. To port these vital capabilities into data-rich broadband networks, 3GPP formalized support for Mission Critical Services (MCX), including voice (MCPTT), video (MCVideo), and data (MCData), beginning in Release 13 and expanding through subsequent specifications \cite{3gpp_ts23280, 3gpp_ts22179}. These frameworks utilize network-embedded Quality of Service (QoS) class prioritization to guarantee deterministic scheduling. However, as independent vendors and operators implement these specifications, ensuring cross-carrier and cross-application compliance remains challenging. Normalizing these implementations requires strict vendor adherence to open 3GPP specifications, stable Interworking Functions (IWF) for legacy LMR-to-broadband translation, and open network interfaces. Scheduled plugtests and conformance evaluations, such as global ETSI events, are critical to empirically validate multi-vendor setups and catch signaling anomalies before active deployment \cite{plugtest25}. ETSI and the United States Department of Homeland Security are pursuing a standards conformance certification program working with the Global Certification Forum (GCF).

\begin{figure*}[htbp]
    \centering
    \includegraphics[width=0.72\textwidth]{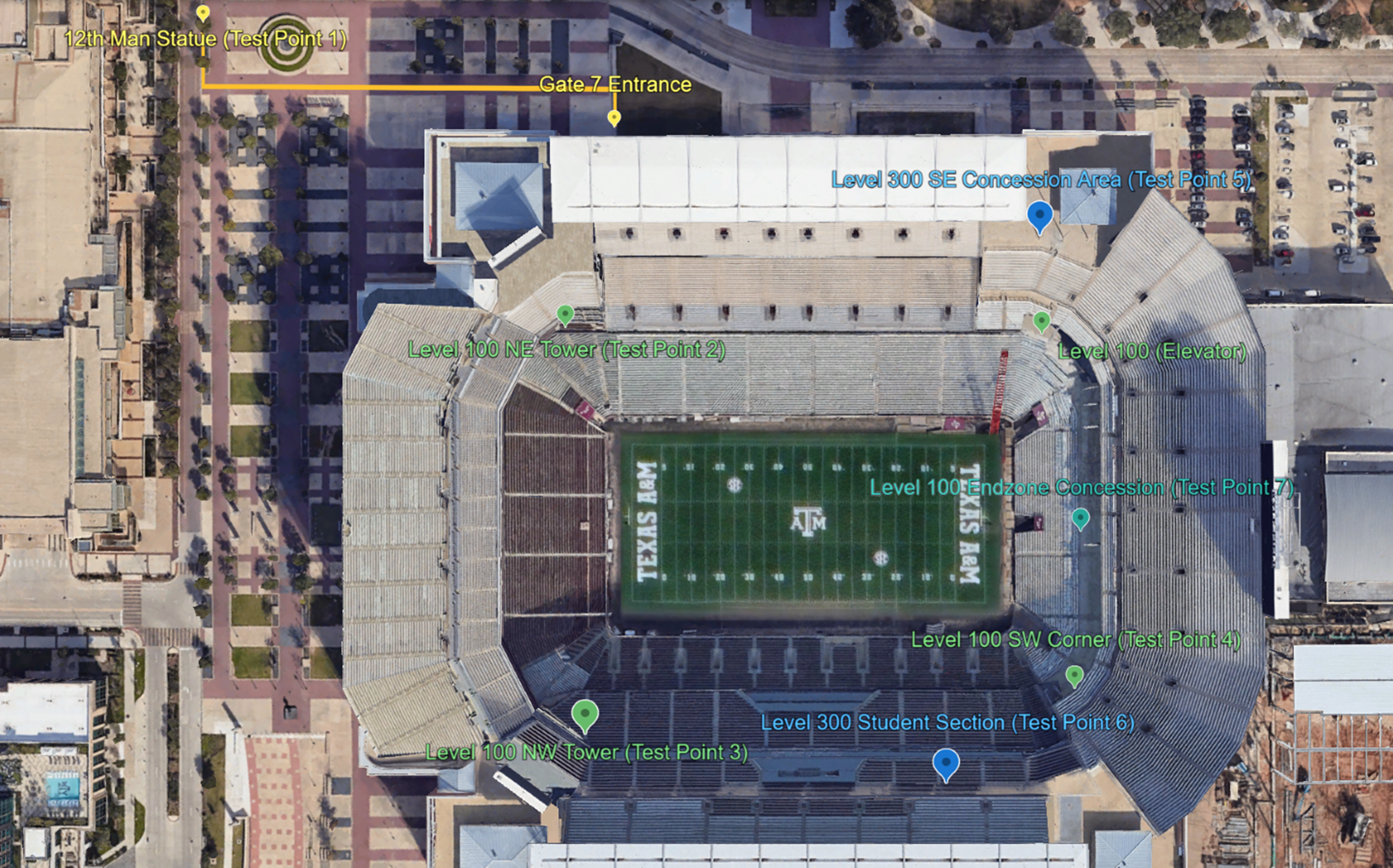}
    \caption{Spatial layout and geographical markers of Test Points (TP 1 to TP 7) at Kyle Field.}
    \label{fig:stadium_map}
\end{figure*}
\section{Experimental Design and Methodology}
\label{sec:methodology}

\begin{figure}[htbp]
    \centering
    \begin{subfigure}[b]{0.26\textwidth}
        \centering
        \includegraphics[width=0.4\textwidth]{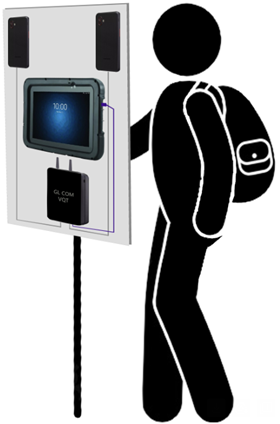}
        \caption{Rig component layout.}
        \label{fig:rig_schematic}
    \end{subfigure}
    \hfill
    \begin{subfigure}[b]{0.29\textwidth}
        \centering
        \includegraphics[width=\textwidth]{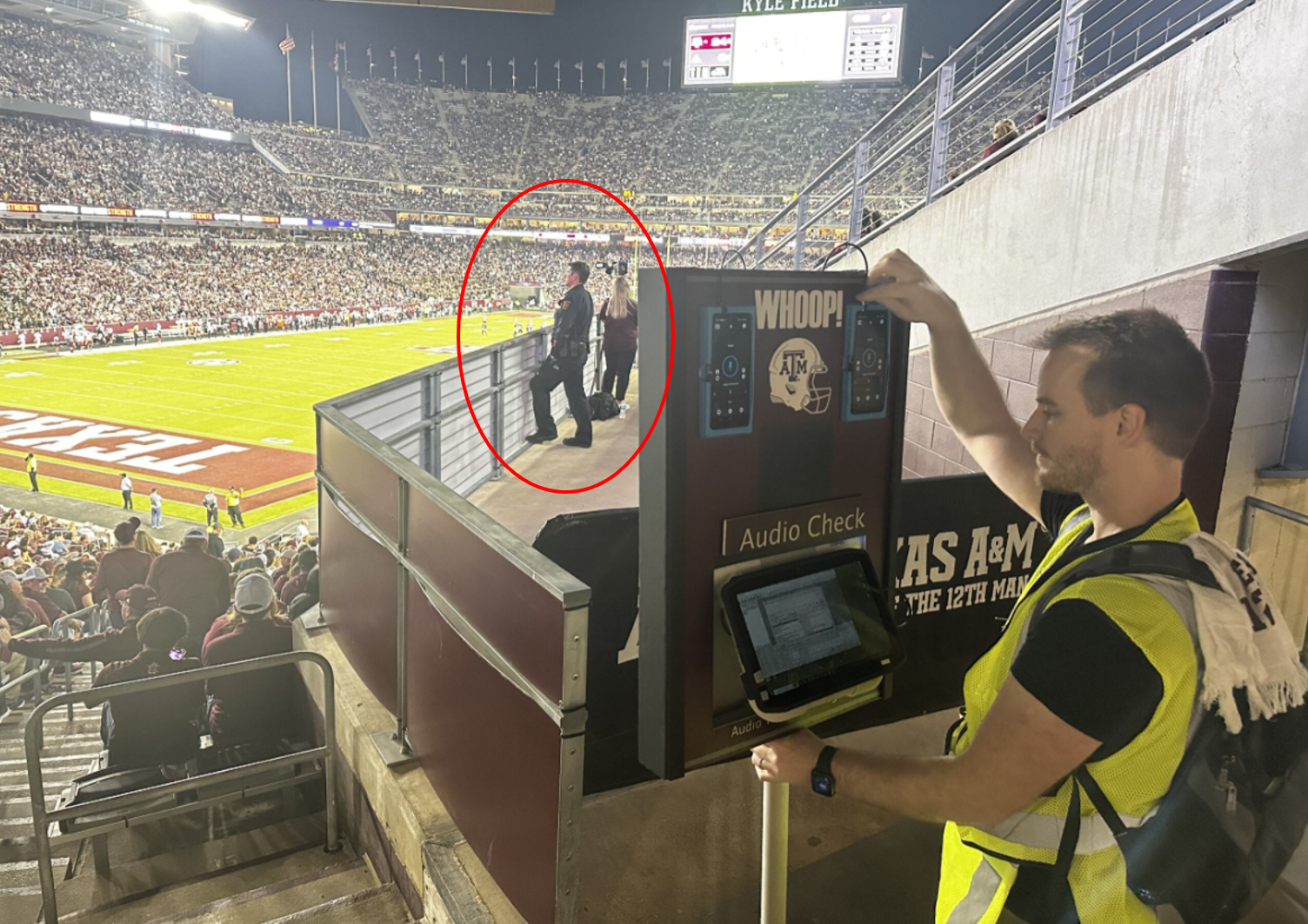}
        \caption{Live field deployment.}
        \label{fig:rig_field}
    \end{subfigure}
    \caption{Multi-carrier testing rig showing hardware concealment within insulated foam board signage, PVC frame scaffolding, and structural phone configuration at head height.}
    \label{fig:test_rig_assembly}
    \vspace{-15pt}
\end{figure}

To empirically evaluate the performance, reliability, and cross-carrier interoperability of public safety communications under load, a field-testing architecture was deployed during an active mass-gathering spectator event. This section details the environmental scenarios, the underlying system configurations, the custom hardware isolation apparatus, and the  metrics used to quantify call observations.

\subsubsection{Communication Scenarios and Spatial Test Points}
The field tests were run at Kyle Field, a sports stadium at Texas A\&M University (College Station, Texas), during a live football game with an active attendance of 105,815 spectators. This mass deployment creates an optimal environment for testing OTT MCX and MCPTT under local network capacity limits and evaluating uplink and downlink resource congestion impacts in performance.

To thoroughly capture network behaviors across disparate RF propagation pathways and infrastructure boundaries, the study was divided into two distinct environmental \textit{scenarios} across seven designated \textit{test points} (TP) as illustrated in Fig.~\ref{fig:stadium_map}:
\begin{itemize}
    \item \textit{Macro Cell Network Scenario (TP 1):} Situated outside the stadium footprint at the 12th Man Statue. Call observations in this zone compete directly with dense tailgating crowds and rely on macro base stations, specifically an off-venue cell site located on Rudder Tower.
    \item \textit{Distributed Antenna System (DAS) Scenario (TP 2--7):} Encompasses six internal venue positions selected because they represent critical operational security posts or experience extreme crowd-driven resource starvation. These positions include: \textit{TP 2 (Level 100 NE Tower):} Highly populated lower concourse. \textit{TP 3 (Level 100 NW Tower):} Concrete structural shielding. \textit{TP 4 (Level 100 SW Corner):} Bowl-adjacent multipath zone. \textit{TP 5 (Level 300 SE Concession Area):} Upper-tier structural attenuation. \textit{TP 6 (Level 300 Student Section):} Peak local user density and continuous uplink signaling load. \textit{TP 7 (Level 100 Endzone Concession):} Heavy crowd bottleneck zone.
\end{itemize}

\begin{figure*}[htbp]
\centering
\includegraphics[width=0.7\textwidth]{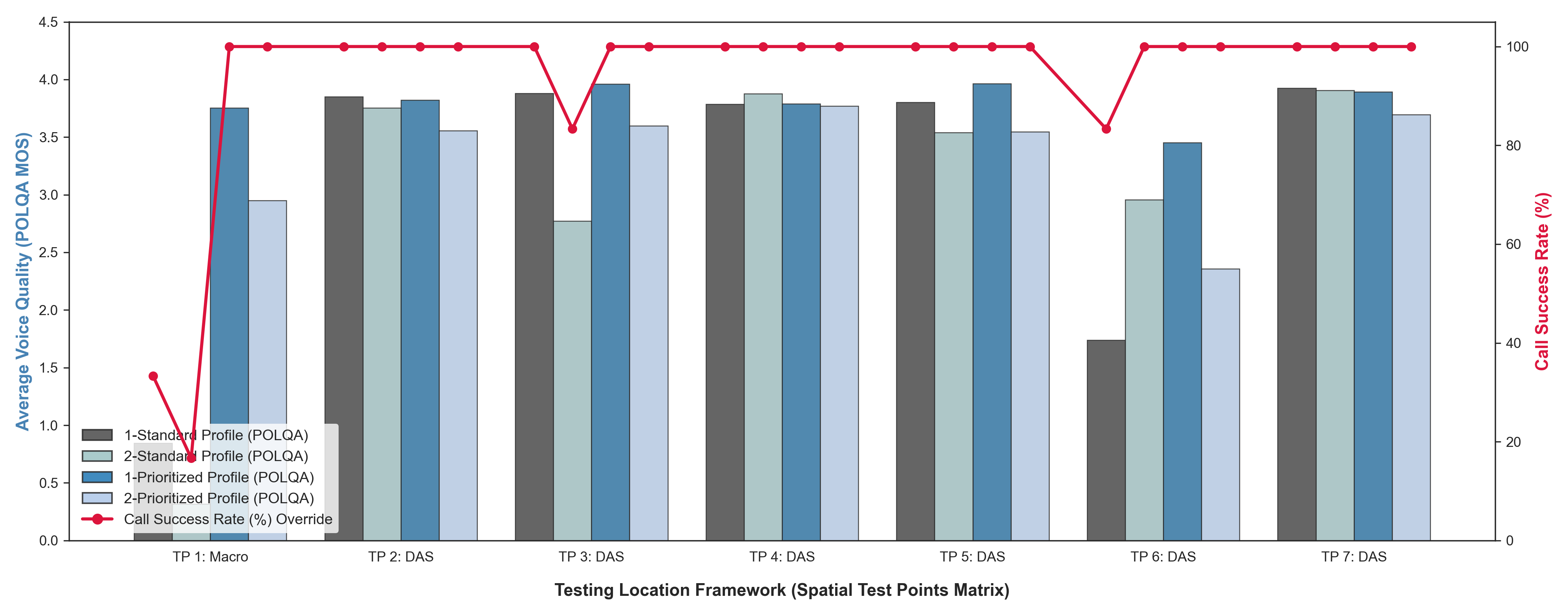}
\caption{Empirical relationship between mean voice quality (POLQA MOS) and call success rates across testing points.}
\label{fig:macro_success}
\vspace{-10pt}
\end{figure*}

\subsubsection{Experimental Configurations}
To isolate carrier network performance without introducing device-specific transceiver variations, a standardized pool of twelve identical Android smartphones was utilized. The devices were provisioned into four distinct testing \textit{configurations} distributed equally across two commercial providers (Carrier 1 and Carrier 2). For each carrier, two active field devices were deployed simultaneously during a run sequence, alongside a single system reserve unit to guarantee testing continuity.

The evaluated network configurations are defined as follows:
\begin{itemize}
    \item \textit{Prioritized Mission-Critical PTT (MCPTT) Configuration:} Configured using carrier-specific variants of the 3GPP-compliant MCX application layer. These clients register directly with dedicated carrier-managed Kodiak server architectures to claim network-layer priority allocation (QoS priority profiles).
    \item \textit{Standard Over-the-Top PTT (OTT PTT) Configuration:} Configured using commercial, non-prioritized best-effort carrier push-to-talk application profiles, operating without underlying resource pre-emption or bearer guarantees.
\end{itemize}

The physical design of the field test rig was engineered using a PVC frame structural skeleton clad inside layers of protective foamboard signage. This apparatus permitted the stable orientation of paired smartphones at approximate head height ($\sim$2 meters off the ground) while masking the tracking electronics from crowd disruption.

\subsubsection{Automation Framework and Call Sample Evaluation}
Automated run sequences were controlled via GL Communications' VQuad™ software coupled to a Dual Universal Telephone Interface (UTI) hardware controller. This managed balanced physical audio lines connected directly to the smartphones' audio interfaces, transmitting a reference audio file into the transmitter chain and capturing the resulting audio from the receiver. At each TP, a continuous loop of automated call sequences was generated, yielding a large dataset of distinct call observations. The captured call samples were parsed through GL Communications' Voice Quality Test (VQT) software to collect five performance metrics:

\textit{Perceptual Speech Quality (POLQA):} Objective auditory degradation per call sample was mapped utilizing the Perceptual Objective Listening Quality Assessment algorithm defined under ITU-T Recommendation P.863. POLQA provides advanced psychoacoustic modeling for wideband HD-speech coding over LTE/5G transport networks, outputting a Mean Opinion Score (MOS) scaling from 1.0 (unacceptable) to 4.5 (excellent). Operationally, any individual call observation that encountered a complete session failure, drop-out, or failed to pass an audio path setup was marked as a critical connection loss and encoded as 0.0 to distinguish complete service failure from low-quality but completed calls.

\textit{Transmission Planning Mapping (E-Model R-Factor): }To bridge perceptual scores with traditional wireless network design criteria, the software calculated an equivalent equipment impairment factor ($I_e$), mapping the POLQA score directly onto the ITU-T G.107 E-Model scale to output an R-Factor (0--100). Values tracking strictly below 60 denote communication links that are unviable for public safety deployment.

\textit{Network Delay: }The end-to-end delay (in milliseconds) required for individual voice utterances to navigate to the receiver. The testing suite captures Minimum, Maximum, and Average Temporal Offsets per call sample to measure transmission lag added by the network.

\textit{Packet Jitter: }Quantifies the statistical variance in arrival intervals between successive received speech utterances (measured in ms) within a single call sample. Jitter values evaluate arrival path volatility, exposing underlying link stability issues under extreme crowd load.


\begin{figure}[htbp]
\centering
\includegraphics[width=0.48\textwidth]{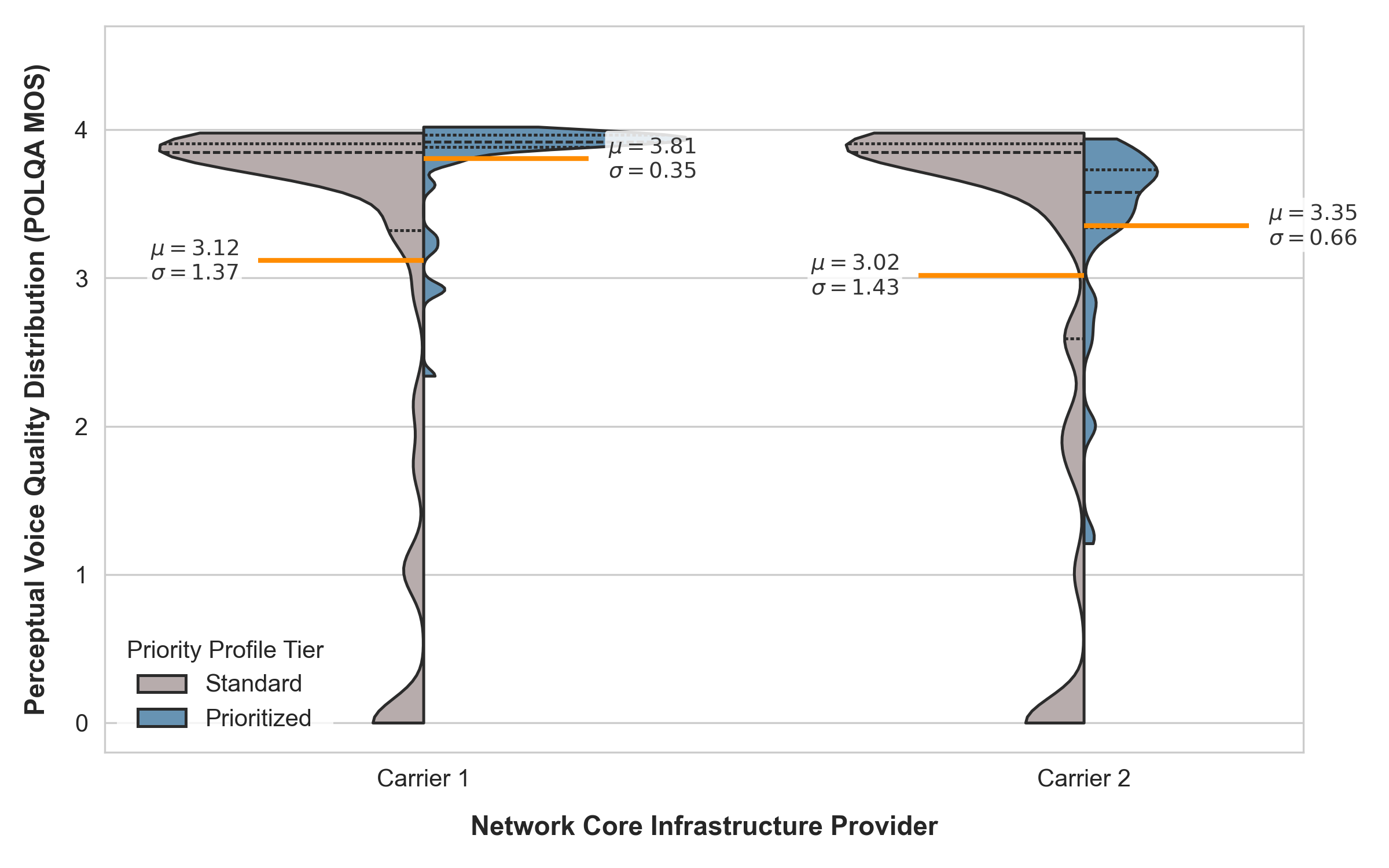}
\caption{Split-violin density profile highlighting empirical voice quality distribution.}
\label{fig:split_density}
\vspace{-10pt}
\end{figure}

\section{Empirical Results and Data Analysis} \label{sec:results}

This section presents the findings gathered from our dense user-crowd stadium deployment. The collected dataset is analyzed to characterize the performance gaps between standard commercial tiers and prioritized network profiles.

\subsection{Voice Quality and Connectivity Reliability} \label{sub:network_perf}

\begin{figure}[htbp!]
\centering
\includegraphics[width=0.48\textwidth]{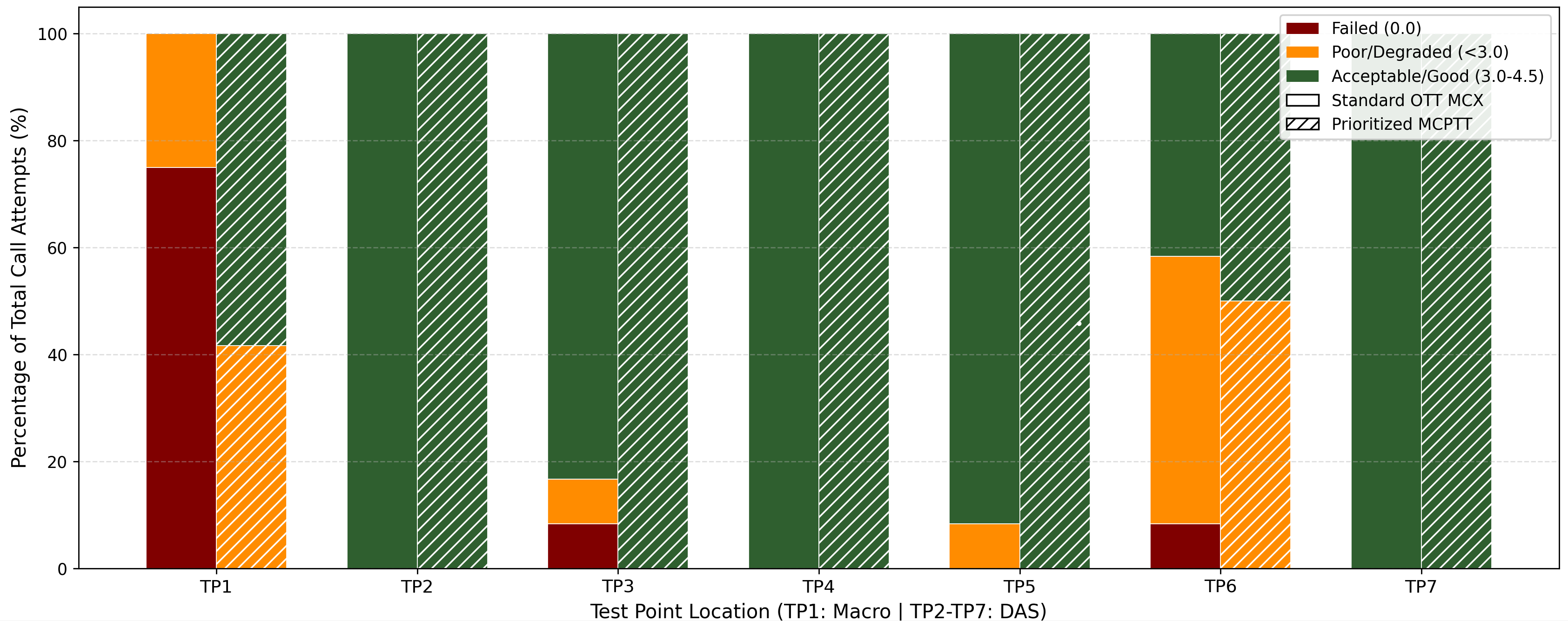}
\caption{Connectivity reliability and quality distribution across macro (TP1) and DAS (TP2-7) testing points.}
\label{fig:connectivity_gap}
\vspace{-10pt}
\end{figure}

Empirical observations across the stadium reveal a reliability gap. Fig.~\ref{fig:macro_success} and the connectivity analysis in Fig.~\ref{fig:connectivity_gap} quantify this divergence. At the Macro-Cell location (TP1), Standard OTT MCX collapsed to a mean POLQA score of 0.58. This failure is primarily driven by connection unavailability (maroon segments in Fig.~\ref{fig:connectivity_gap}) rather than incremental audio degradation. Prioritized MCPTT maintained a MOS of 3.28 at TP1, bypassing macro-cell resource block (RB) starvation through standardized priority signaling. Inside the DAS environment (TP2--TP7), overall reliability improved; however, localized saturation in the Student Section (TP6) induced structural degradation. Even with prioritization, POLQA scores at TP6 dropped to a mean of 2.91, indicating that extreme uplink signaling loads can challenge secured resource allocations. Fig.~\ref{fig:split_density} illustrates this via a split-violin density matrix. Standard tiers exhibit unstable, multi-modal distributions with wide variance ($\sigma = 1.10$ to $1.43$), while prioritized profiles concentrate density into sharp unimodal peaks (3.8--4.2 MOS) with significantly tighter variance ($\sigma = 0.35$ for Carrier 1).

\subsection{Transport-Layer Failure Boundaries} \label{sub:transport_cliff}

\begin{figure}[htbp]
\centering
\includegraphics[width=0.4\textwidth]{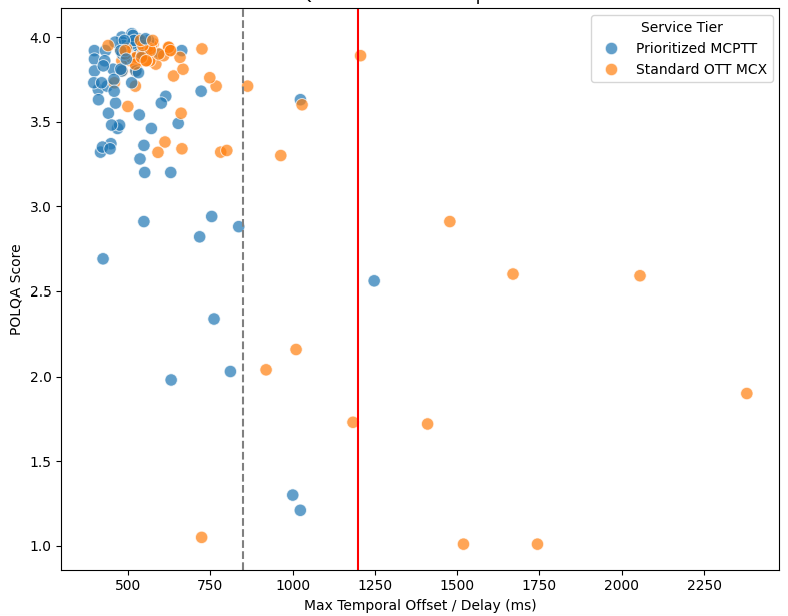}
\caption{POLQA vs. Maximum Temporal Offset highlighting a 1200 ms Operational Drop Boundary.}
\label{fig:polqa_delay}
\vspace{-10pt}
\end{figure}

\begin{figure}[htbp]
\centering
\includegraphics[width=0.4\textwidth]{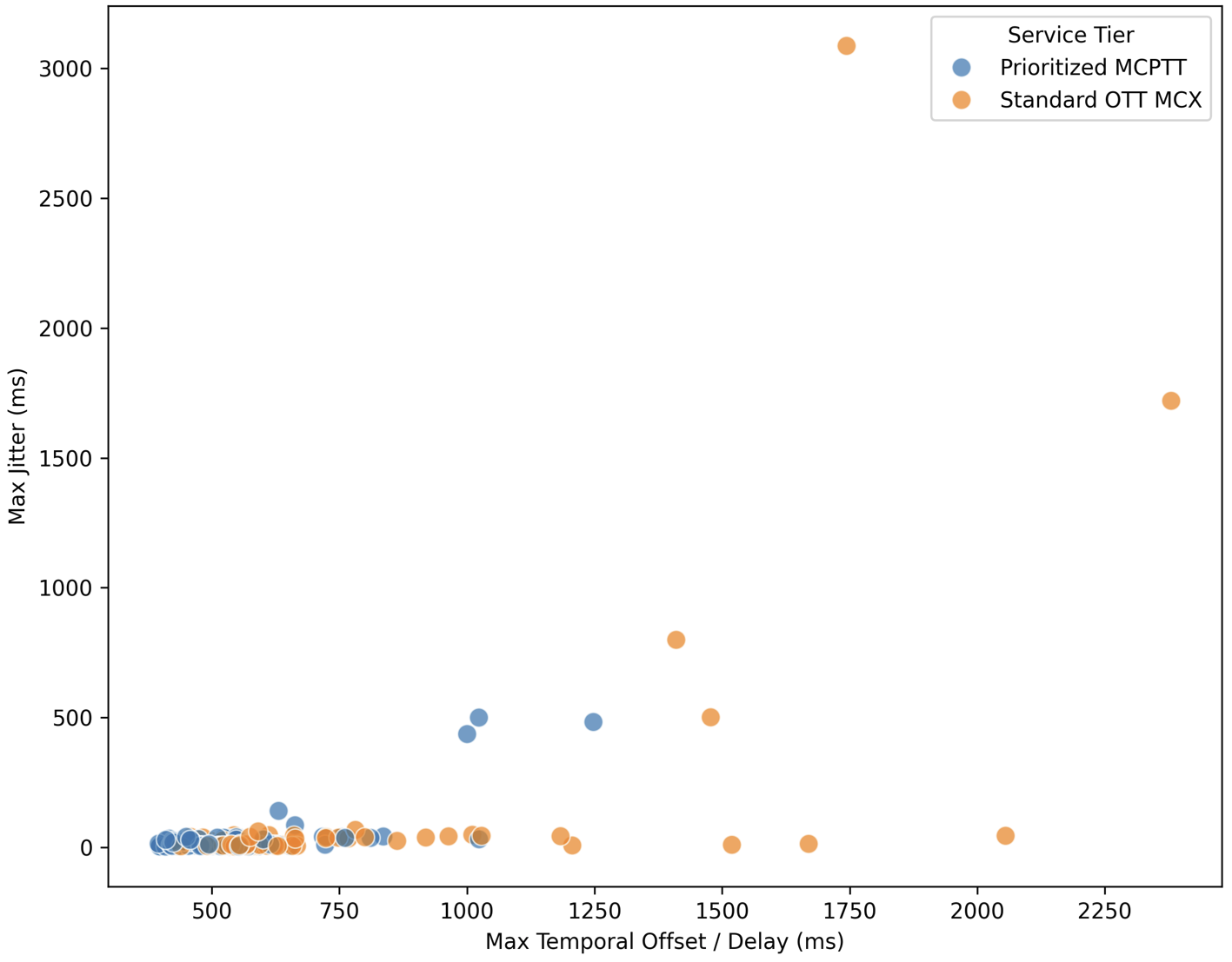}
\caption{Maximum Jitter vs. Maximum Delay.}
\label{fig:jitter_delay}
\vspace{-10pt}
\end{figure}

To identify infrastructure thresholds triggering voice path collapse, Fig.~\ref{fig:polqa_delay} correlates perceptual quality with transport-layer stress. Audio quality remains stable until Maximum Temporal Offset (Delay) reaches a degradation Knee at 850 ms. Beyond the Operational Drop Boundary of 1200 ms, packet delivery variations overwhelm adaptive de-jitter buffers, probably causing codec collapse. Fig.~\ref{fig:jitter_delay} tracks the interaction between Maximum Jitter and Delay. Structural failure occurs when jitter exceed 150 ms, triggering consecutive frame drops and forcing POLQA scores to bottom out. These findings highlight that prioritizing the bearer can support transport-layer metrics bounded away from failures thresholds.

\subsection{Terminal Equipment Invariance} \label{sub:hardware_invariance}
\begin{figure}[htbp!]
\centering
\includegraphics[width=0.45\textwidth]{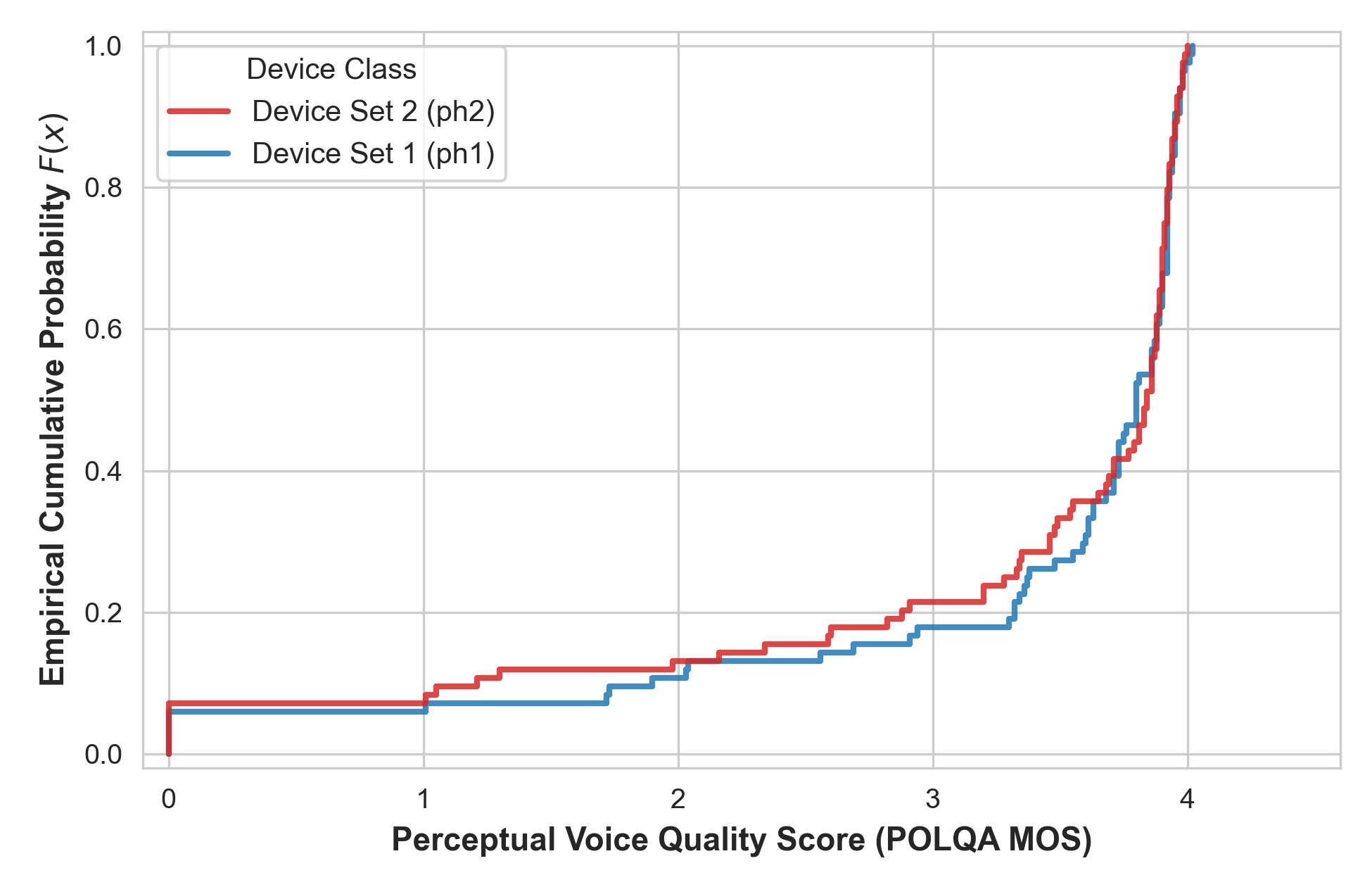}
\caption{Global ECDF showing performance equivalence and statistical invariance between testing handsets.}
\label{fig:invariance_ecdf}
\vspace{-10pt}
\end{figure}

\begin{figure}[htbp]
\centering
\includegraphics[width=0.48\textwidth]{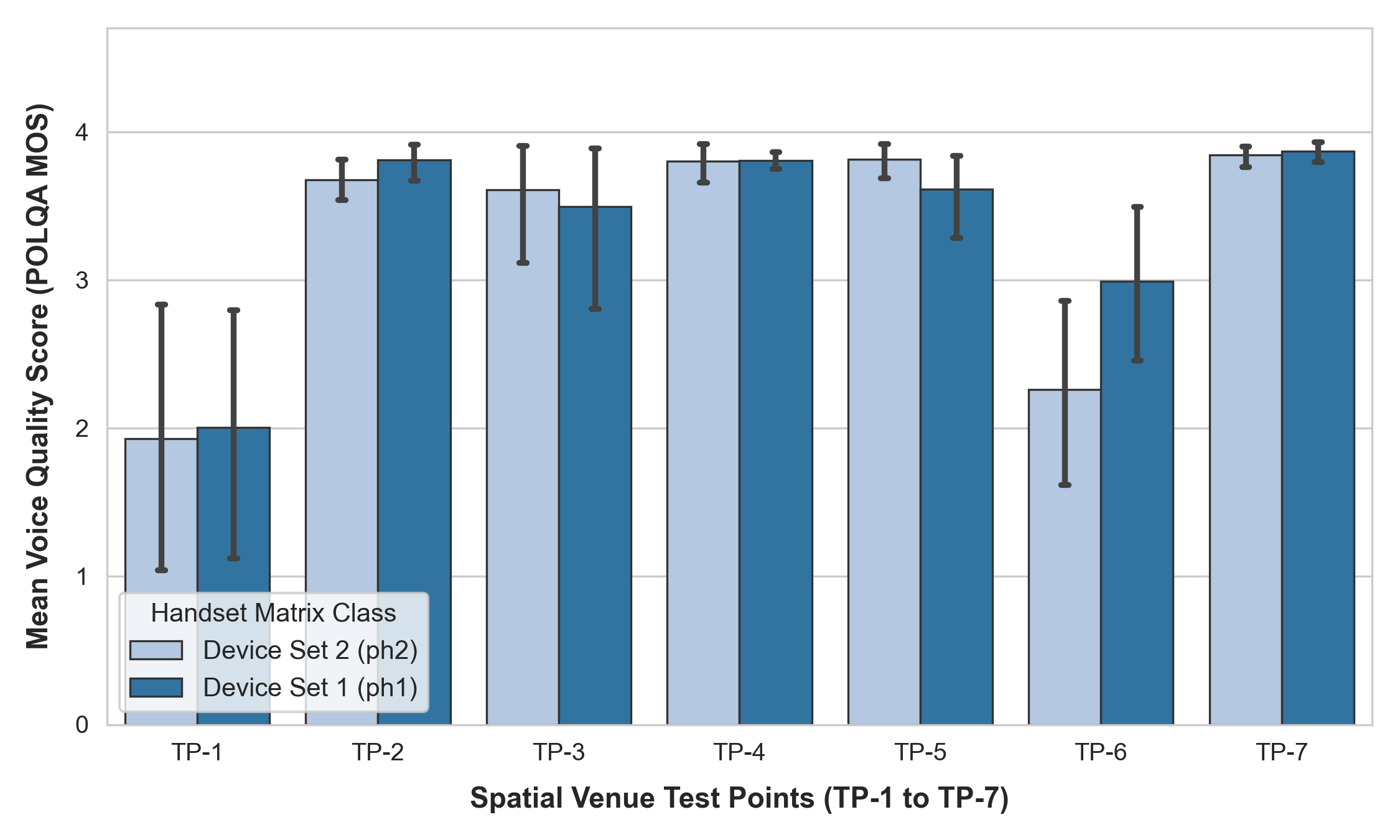}
\caption{Mean POLQA MOS across Test Points 1 through 7 for both handsets confirming uniform environmental impact.}
\label{fig:invariance_spatial}
\vspace{-10pt}
\end{figure}

To ensure findings reflect network infrastructure limits rather than hardware bias, we cross-examined the two identical handsets ($ph1$, $ph2$). Fig.~\ref{fig:invariance_ecdf} shows a nearly identical global Empirical Cumulative Distribution Function (ECDF) lines. Parametric ANOVA yields an $F$-statistic of 0.3149 ($p = 0.5754$), failing to reject the null hypothesis of hardware bias. Further profiling in Fig.~\ref{fig:invariance_spatial} confirm that mean performance bars and confidence intervals overlap across all test points and configuration tiers. This statistical invariance confirms that documented degradations is likely driven by RAN resource exhaustion and core queuing policies.

\section{Discussion and Operational Insights}
\label{sec:discussion}

\subsubsection{Prioritized Interoperable Profiles}
\label{sub:quantifiable_advantages}

This empirical study identifies critical failure states in standard best-effort OTT applications under peak network load, where resource block (RB) starvation induces unstable, multi-modal POLQA distributions with mean scores falling beneath 2.5~MOS and connection success rates collapsing to nearly 60\%. Conversely, 3GPP-compliant prioritization (MCPTT) mitigates these operational risks by securing fixed Allocation and Retention Priority (ARP) and explicit QoS Class Identifier (QCI) profiles to bypass standard MAC-layer scheduling bottlenecks. This structural isolation tightens voice path variance ($\sigma$ tracking as low as 0.35) into a stable unimodal peak, maintaining consistent performance between 3.4 and 4.1~POLQA MOS with a near-100\% success rate, ensuring mission-critical reliability during severe infrastructure strain.

\subsubsection{Cross-Carrier Interoperability}
\label{sub:interoperability_bottlenecks}

Evaluating multi-carrier performance reveals that While prioritization improves results for all providers, baseline standard traffic exhibits noticeable variance: Carrier 1 yields $\mu = 3.16$ ($\sigma = 1.10$), whereas Carrier 2 drops to $\mu = 2.97$ under identical crowd loads. These gaps are likely driven by carrier-specific Distributed Antenna System (DAS) sector boundaries and scheduling loops. Architectural constraints exacerbate these vulnerabilities; at concrete-shadowed locations like the Level 100 SW Corner (TP4) and Level 300 SE Corner (TP5), multipath interference combines with congestion to push maximum jitter beyond the 150~ms "packet-delivery cliff edge". As mapped in Fig.~\ref{fig:polqa_delay} and Fig.~\ref{fig:jitter_delay} how operating beyond this value overwhelms buffers and induces codec state machine collapse, causing voice paths to fail. These results demonstrate that cross-carrier interoperability requires strict standardization of both radio priority layers and core transport-layer queuing policies to prevent boundary-induced link failures.

\subsubsection{First Responders and Mission-Critical Planning}
\label{sub:operational_implications}
Mapping these findings into operational guidelines provides insights for joint-agency public safety planning. First, because voice path collapse follows a non-linear cliff-edge model driven by infrastructure queuing, planners should not rely on standard statistical averages such as mean delay or average packet loss to gauge system reliability. System health can be inferred by the upper tail of the jitter distribution. Hence, mission planning must mandate continuous end-to-end network slicing and/or priority mappings to ensure critical paths remain bounded away from this failure boundary~\cite{chowdhury2025framework}. Second, validation using identical handsets confirms performance bottlenecks in high-density environments should focus on addressing the constraint existing within the network infrastructure, public safety entities must rely on pre-negotiated Mutual Aid Roaming agreements and hard-coded cross-carrier QoS priority definitions to ensure uniform, prioritized channel access before transport queues saturate.

\section{Conclusion and Future Work}
\label{sec:conclusion}
The empirical results of this study show that MCPTT degradation under extreme density stems from network infrastructure specifically Distributed Antenna System (DAS) schedulers and backhaul queues. Voice path performance follows a non-linear failure model: once transport jitter breaches a critical threshold, adaptive handset buffers experience systematic underflows, breaking packet continuity and triggering immediate call failure. Conversely, Quality-of-Service (QoS) tiering appears to reduce exposure to standard congestion queues to keep jitter safely below this threshold. Thus, guaranteeing link survivability during mass-crowd events requires end-to-end network slicing, explicit radio resource reservation, and hard-coded cross-carrier priority mappings over physical densification. Future work should evaluate these prioritization models within Next Gen (e.g. standalone 5G topologies), investigate cross-carrier handover latency penalties, analyze Multi-access Edge Computing (MEC) mitigation of backhaul queuing, and integrate real-time mission-critical video and data telemetry streams.

\section*{Acknowledgment}
This material is based upon work supported by the U.S. Department of Homeland Security under Grant Award Number 70RSAT21G00000012. Disclaimer: “The views and conclusions contained in this document are those of the authors and should not be interpreted as necessarily representing the official policies, either expressed or implied, of the U.S. Department of Homeland Security. Gemini AI tools were employed in refining the
manuscript technical narrative. 

\bibliographystyle{IEEEtran}
\bibliography{references}

\end{document}